\documentstyle[epsfig]{mn}

\title[Recent solar neighbourhood star formation history]{The recent star
       formation history of the Hipparcos solar neighbourhood}

\author[X. Hernandez, D. Valls-Gabaud and G. Gilmore]{
X. Hernandez$^1$, David Valls-Gabaud$^{2}$ and Gerard Gilmore$^3$\\
$^1$ Osservatorio Astrofisico di Arcetri, Largo E. Fermi 5, 
	50125 Firenze, Italy\\
$^2$ Laboratoire d'Astrophysique, UMR CNRS 5572, 
	Observatoire Midi-Pyr\'en\'ees, 14 Avenue E. Belin, 
	31400 Toulouse, France\\
$^3$ Institute of Astronomy, Cambridge University, 
	Madingley Road, Cambridge CB3 0HA, UK \\
}

\date{\today}

\begin{document}
\maketitle

\begin{abstract}

We use data from the Hipparcos catalogue to construct colour-magnitude 
diagrams for the 
solar neighbourhood, which are then treated using advanced Bayesian analysis
techniques to derive the star formation history, $SFR(t)$, of this region 
over the last 3 Gyr. The method we use allows the recovery of the underlying
 $SFR(t)$ without the need of assuming any {\it a priori} structure or 
condition on $SFR(t)$, and hence yields a highly objective result. The 
remarkable accuracy of the data permits the reconstruction of the local 
$SFR(t)$ with an unprecedented time resolution of $\approx 50 $ Myr. 
A $SFR(t)$ having an oscillatory component of period $\approx 0.5$ Gyr is 
found, superimposed on a small level of constant star formation activity.
Problems arising from the non-uniform selection function of the 
Hipparcos satellite are discussed and treated. Detailed statistical tests 
are then performed on the results, which confirm the inferred $SFR(t)$ 
to be compatible with the observed distribution of stars.

\end{abstract}


\section{Introduction}

The problem of deducing the star formation rate history, $SFR(t)$, of 
the Milky Way has been generally  attempted in terms of indirect inferences
mostly through chemical evolution models. 
The validity of these methods relies on the soundness of the assignation of
a ``chemical age'' to each of the studied stars. Generally a metallicity
indicator is chosen, and used to measure the metal content of a number of stars
which are then binned into age groups through the use of an age-metallicity
relation derived from a chemical evolution model. For example, Rocha-Pinto \&
Maciel (1997) take a variety of age-metallicity relations (AMRs) from the 
literature and use a closed box chemical evolution model to translate AMRs 
into $SFR(t)s$, allowing for an intrinsic Gaussian spread in the AMR assumed 
to be constant in time. 

The advantages of these methods are that large samples of stars both in the 
solar neighbourhood  and further away within the disk of the galaxy can be 
studied. An inferred $SFR(t)$ can be constructed over an ample time range 
and spatial extent within the Galaxy which is consistent with the 
metallicities of the sample studied, and the chemical evolution model proposed.
However, the validity of the AMR assumption can not be checked independently 
of the proposed chemical evolution model, and is necessarily dependent on 
what is chosen for the mixing physics of the ISM, the possible infall of 
primordial non enriched gas,  and the still largely unknown galactic 
formation scenario in general. This last problem also affects attempts at 
inferring the $SFR(t)$ from stellar kinematics (e.g. Gomez et al. 1990, 
Marsakov et al., 1990).

With the recent availability of the Hipparcos satellite catalogue (ESA 1997)
we are now in a position to attempt recovery of the local $SFR(t)$ directly,
without the need of any model dependent assumptions. Previous direct approaches have been undertaken through the binning of observed stars into age groups 
according to the degree  of chromospheric activity as measured through 
selected emission line features, with conflicting results depending on the
assumed age-activity relation (e.g. Barry, 1988, and Soderblom et al., 1991).
Using this technique, Rocha-Pinto et al. (1999) have
derived a star formation history from the chromospheric activity-age 
distribution of a larger sample comprising 552 stars, founding evidence for
intermittency in the $SFR(t)$ over 14 Gyr.
The Hipparcos catalogue offers high quality photometric data for a large 
number of stars in the solar neighbourhood, which can be used to construct a 
colour-magnitude diagram (CMD) for this region. Once a CMD is available,
it is in principle possible to recover the $SFR(t)$ which gave rise to the 
observed distribution of stars, assuming only a stellar evolutionary model in 
terms of a set 
of stellar tracks. In practice the most common approach to inverting CMDs has
been to propose a certain parameterization for the $SFR(t)$, which is used to
construct synthetic CMDs, which are statistically compared to the observed 
ones to select the values of the parameters which result in a best match CMD.
Examples of the above are  Chiosi et al. (1989), 
Aparicio et al. (1990) and Mould et al. (1997) using Magellanic
and local star clusters, and Mighell \& Butcher (1992), Smecker-Hane et al. 
(1994),  Tolstoy \& Saha (1996), Aparicio \& Gallart (1995) and Mighell 
(1997) using local dSph's.

We have extended these methods in Hernandez et. al. (1999) (henceforth 
paper I) by combining a rigorous maximum likelihood statistical approach, 
analogous to what was introduced by Tolstoy \& Saha (1996), with a variational
calculus treatment. This  allows a totally non-parametric solution of the
problem, where no {\it a priori} assumptions are introduced. This method was
applied by Hernandez et al. (2000) (paper II) to a set of HST CMDs of local
dSph galaxies to infer the $SFR(t)$ of these interesting systems. The result
differs from what can be obtained from a chemical evolution model in that
a direct answer is available, with a time resolution which depends only on 
the accuracy of the observations.

Limitations on the applicability of our method to the Hipparcos data appear
in connection to the selection function of the catalogue. The need to work 
only with complete volume-limited samples limits the age range over which we
can recover the $SFR(t)$ to 0--3 Gyr, with a resolution of $\sim 0.05 $ Gyr.
This makes it impossible to compare our results with those of chemical 
evolution models which typically sample ages of 0--14 Gyr, with resolutions
of 0.5--1.5 Gyr.

In Section 2 we give a summarized review of the method introduced in paper I,
the sample  selection and results are discussed in section 3. Section 4 
presents a careful  statistical testing of our results, and Section 5 our
conclusions.

\section{The method}

In this section we give a summary description of our HR diagram inversion 
method, which was described extensively in our paper I. In contrast with 
other statistical methods, we do not need to construct synthetic colour 
magnitude diagrams for each of the possible  star formation histories being
considered. Rather we use a direct approach which solves  for the best 
$SFR(t)$ compatible with the stellar evolutionary models assumed and the 
observations used. 

The evolutionary model consists of an isochrone library,
and an IMF. Our results are largely insensitive to the details of the 
latter, for which  we use:

\begin{equation}
\rho(m) \propto \left\{
\begin{array}{rl}
     m^{-1.3} &  0.08M_{\odot} <m\le 0.5 M_{\odot} \\[1.0 ex] 
     m^{-2.2} &  0.5M_{\odot} <m\le 1.0 M_{\odot} \\[1.0 ex]
     m^{-2.7} &  1.0M_{\odot} < m
\end{array}\right.
\end{equation} 

The above fit was derived by Kroupa et al. (1993) for a large sample towards
both galactic poles and all the solar neighbourhood, and therefore applies 
to the Hipparcos data.

As we shall be treating here only data from the solar neighbourhood derived
by the Hipparcos satellite, we shall assume for the observed stars a fixed 
metallicity of $[Fe/H]=0$. This assumption is valid as we will only be 
treating stars within a short distance from the Sun, having a small spread 
in ages. Once the metallicity is fixed we use the latest Padova isochrones 
(Fagotto et al., 1994, Girardi et al., 1996) together with a detailed constant
phase interpolation scheme using only stars at constant evolutionary phase,
to construct an isochrone library having a chosen temporal resolution. 

To transform the isochrones from the theoretical HR diagram to
the observed colour-magnitude diagrams, we used the transformations
provided by the calibrations of Lejeune et al. (1997)
which are appropriate for the solar metallicities considered here.
Using the updated calibrations given by Bessell et al. (1998)
does not change the transformations significantly in the regime
used here, unlike the case for giant and AGB stars (see Weiss \&
Salaris, 1999).

In this case we implement the method with a formal resolution of 15 Myr, 
compatible with  the high resolution of the Hipparcos observations. It is
one of the advantages of the method  that this resolution can be increased 
arbitrarily (up to the stellar model resolution) with computation times 
scaling only linearly with it. 

Our only other inputs are the positions of, say, $n$ observed stars in the 
HR diagram, each  having a colour and luminosity, $c_{i}$ and $l_{i}$. 
Starting from a full likelihood model, we first construct the probability 
that the $n$ observed stars resulted from a certain  $SFR(t)$. This will 
be given by:

\begin{equation}
{\cal L}= \prod_{i=1}^{n} \left( 
\int_{t_0} ^{t_1} \, SFR(t) \, G_{i}(t) \, dt \right),
\end{equation}

where

\begin{eqnarray*}
G_{i}(t) & = & \int_{m_0}^{m_1} {\rho(m;t) \over{2 \pi \sigma(l_i) \sigma(c_i)}} \, \times  \\
\lefteqn{\times  \exp\left(-D(l_{i};t,m)^2 \over {2 \sigma^2(l_i)} \right) 
\exp\left(-D(c_{i};t,m)^2 \over {2 \sigma^2(c_i)} \right) \hspace{5pt} 
 dm} \\
\end{eqnarray*}

\begin{figure*}
\epsfig{file=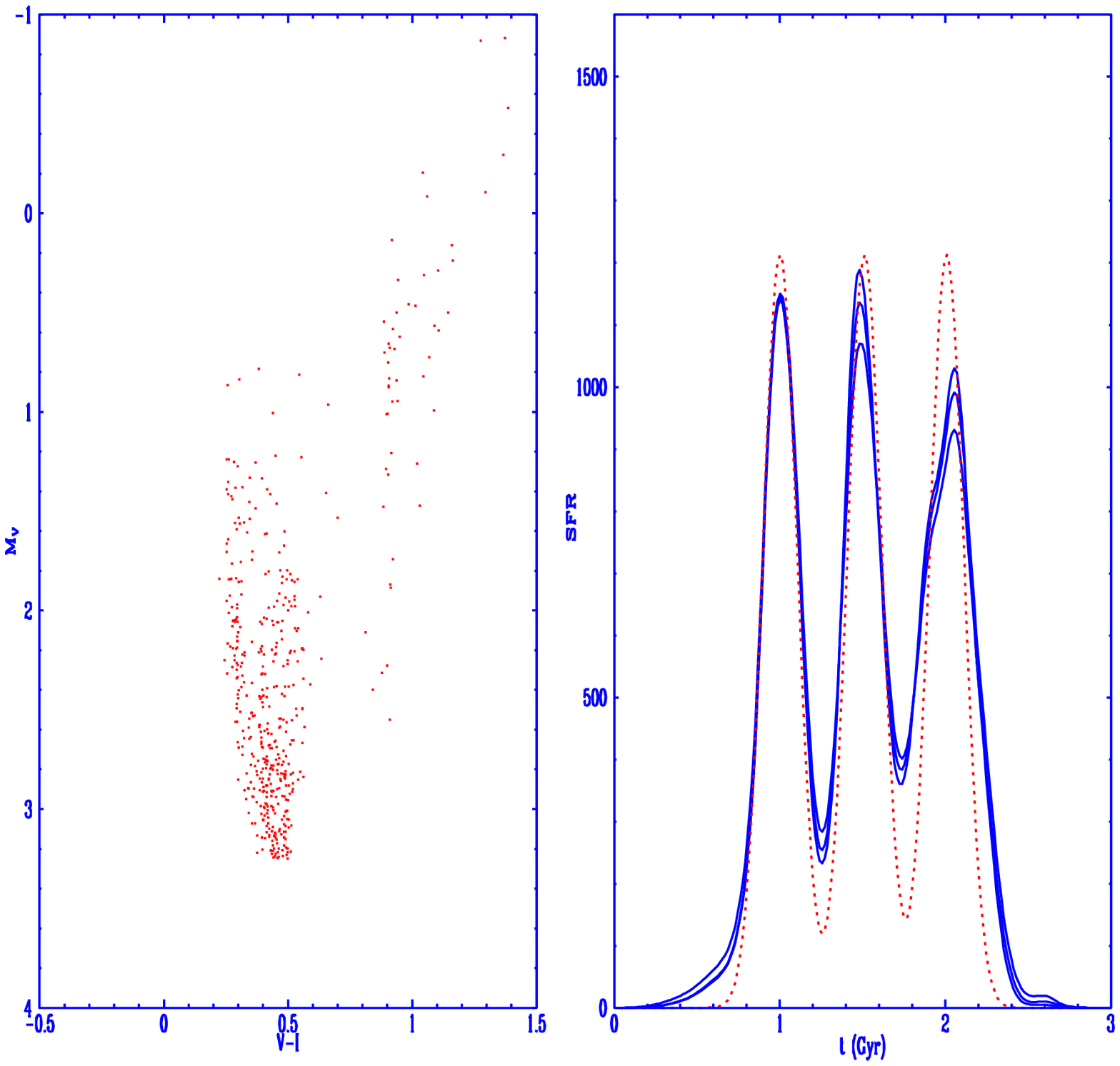,angle=0,width=18.1cm,height=9.0cm}
\@ \textbf{Figure 1.}\hspace{5pt}{
\begin{flushleft}\textbf{Left:} Simulated colour-magnitude diagram resulting
 from the first test SFR(t).
\textbf{Right:} First test SFR(t), dashed line. The solid curves are the 
last 3 iterations of the method, showing an accurate reconstruction of the
 input SFR(t).
\end{flushleft}}
\end{figure*} 

\begin{figure*}
\epsfig{file=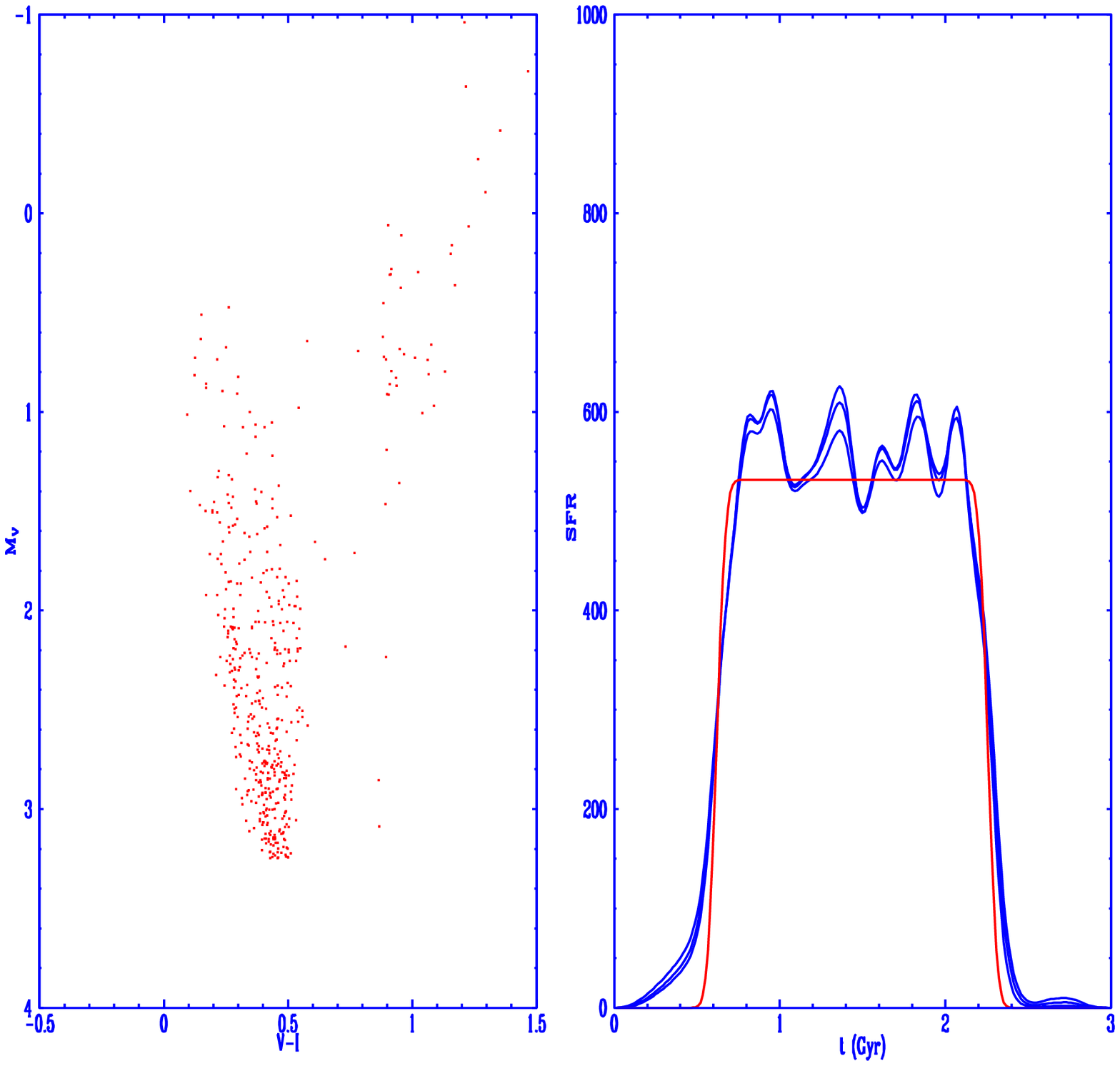,angle=0,width=18.1cm,height=9.0cm}
\@ \textbf{Figure 2.}\hspace{5pt}{
\begin{flushleft}\textbf{Left:} Simulated CMD  resulting from the second
 test SFR(t).
\textbf{Right:} Second test SFR(t), dashed line. The solid curves are the 
last 3 iterations of the method, showing an accurate reconstruction of the
input SFR(t).
\end{flushleft}}
\end{figure*} 

In the above expression $\rho(m;t)$ is the density of points along the 
isochrone of age $t$, around the star of mass $m$, and  is determined by the
assumed IMF together with the duration of the differential phase around the
star of mass $m$. The ages $t_0$ and $t_1$ are a minimum and a maximum age 
needed to be considered, as $m_0$ and $m_1$ are a minimum and a maximum mass
considered along each isochrone, e.g. 0.6 and 20 $M_{\odot}$. $\sigma(l_i)$
and  $\sigma(c_i)$ are the amplitudes of the  observational errors in the 
luminosity and colour of the $i$th star. These values are supplied by the 
particular observational sample one is analysing. Note that for the
sample we have selected (Section 3), there is no correlation between these
errors.  Finally, $D(l_{i};t,m)$ 
and $D(c_{i};t,m)$ are the differences in luminosity and colour, respectively,
between the $i$th observed star and a general star of age and mass $(m,t)$.
We shall refer to $G_{i}(t)$ as the likelihood matrix, since each element
represents the probability that a given star, $i$, was actually formed at
time $t$ with any mass.  
 
A similar version was introduced in paper I, but restricted to the case of 
observational errors only in one variable, which was adequate to the problem
of studying the $SFR(t)$ of local dSph galaxies treated in paper II.
Equation (2) is essentially the extension from the case of a discretised 
$SFR(t)$ used by Tolstoy \& Saha (1996), to the case of a continuous function
in the construction of  the likelihood.  The challenge now is to find the 
optimum $SFR(t)$ without evaluating equation (2) i.e. without introducing 
a fixed set of test $SFR(t)$ cases from which one is selected. 
 
The condition that ${\cal L}(SFR)$ has an extremal can be written as
$$
\delta {\cal L}(SFR)=0,
$$
and a variational calculus treatment of the problem applied.  Firstly, we 
develop the product over $i$ using the chain rule for the variational
derivative, and divide the resulting sum by ${\cal L}$ to obtain:

\begin{equation}
\sum_{i=1}^{n} \left(
{\delta \int_{t_0} ^{t_1} SFR(t) \, G_{i}(t) \, dt} \over {\int_{t_0}^{t_1}
 SFR(t) \, G_{i}(t) \, dt} \right) =0
\end{equation}

Introducing the new variable $Y(t)$ defined as:

$$
Y(t)=\int{ \sqrt {SFR(t)} \, dt}\;  \Longrightarrow \;  SFR(t)=\left( {dY(t)
\over dt} \right)^2
$$

and introducing the above expression into equation~(3) we can develop the
Euler equation to yield 

\begin{equation}
{d^2 Y(t)\over dt^2}\sum_{i=1}^{n} \left( G_{i}(t) \over I(i)\right)
=-{dY(t)\over dt}\sum_{i=1}^{n} \left( dG_{i}/dt \over I(i)\right)
\end{equation}

where 
$$
I(i)=\int_{t_0}^{t_1} SFR(t)\,  G_{i}(t) \, dt
$$

This in effect has transformed the problem from one of searching for a 
function which maximizes a product of integrals (equation 2) to one of 
solving an integro-differential equation (equation 4).  We solve this 
equation iteratively, with the boundary condition $SFR(t_{1})=0$. 

Details of the numerical procedure required to 
ensure convergence to the maximum likelihood $SFR(t)$ can be found in our 
paper I, where a more complete development of the method is also found.
In paper I the method was tested extensively using synthetic HR diagrams,
obtaining very satisfactory results. Equation (4) will be satisfied by any 
stationary point in the likelihood, not just the global maximum, it is therefore
important to check that the numerical algorithm implemented converges to the
true $SFR(t)$. This was shown in our paper I, using synthetic HR diagrams
extensively, and testing the robustness of the method to changes in the initialization
condition of the algorithm. An independent test of the validity of the results was also
implemented, and is described in section 4.

The main advantages of our method over
other maximum likelihood schemes are (1) the totally non parametric approach 
the variational calculus treatment allows, and (2) the efficient computational
procedure, where no time consuming repeated comparisons between synthetic 
and observational  CMD are necessary, as the optimal $SFR(t)$ is solved 
for directly.

\subsection{Two tests}

We now present two examples of the method's performance, in cases similar 
to the Hipparcos  samples CMDs. The left panel of Figure (1) shows a 
synthetic CMD produced from the first input $SFR(t)$, resulting in a number
of stars similar to what the Hipparcos samples yield for small errors in 
$V-I$ ($<0.12$ mag) and $M_{V}$ $(<0.02$ mag). The positions of the simulated
stars are then used to construct the likelihood matrix,which is used to 
recover the inferred $SFR(t)$, through an iterative numerical procedure 
(see paper I). The right panel of Figure (1) shows the last 3 iterations of 
the method (solid curves) and the input $SFR(t)$, a three burst $SFR(t)$  
(dotted curve). It can be seen that the main features of the input $SFR(t)$ 
are accurately recovered. The age, duration and shape of the input $SFR(t)$
are clearly well represented by the final inferred  $SFR(t)$. As the 
difference between successive isochrones diminishes with age, since the 
errors remain constant, the accuracy of the recovery procedure diminishes 
with the age of the stellar populations being treated. This is seen in that
the first burst is very accurately recovered, whilst the last one appears 
somewhat spread out. 

The last example is shown in Figure (2), which is analogous to Figure (1). 
Here a $SFR(t)$ which is  constant over a large period is treated. The HR 
diagram of this case appears by sight almost identical to that of the 
previous case, however, given the extremely small errors assumed (typical of
the Hipparcos data) the method is capable of distinguish and accurately 
recover the input $SFR(t)$ of these two cases. The small number of stars 
$(\sim 450)$ result in a degree of shot noise, which has to be  artificially
suppressed using a smoothing procedure, the result of which is seen in the 
residual short period oscillations of the inferred $SFR(t)$. This smoothing
procedure reduces the effective resolution of the method to $50$ Myr. Note 
that as in 
the previous example, the inversion method successfully recovers the main 
features of the input $SFR(t)$. In these two tests only stars  bluewards 
of $V-I=0.7$ where considered in the inversion procedure (see below).

\section{Sample selection and results}

\begin{figure}
\epsfig{file=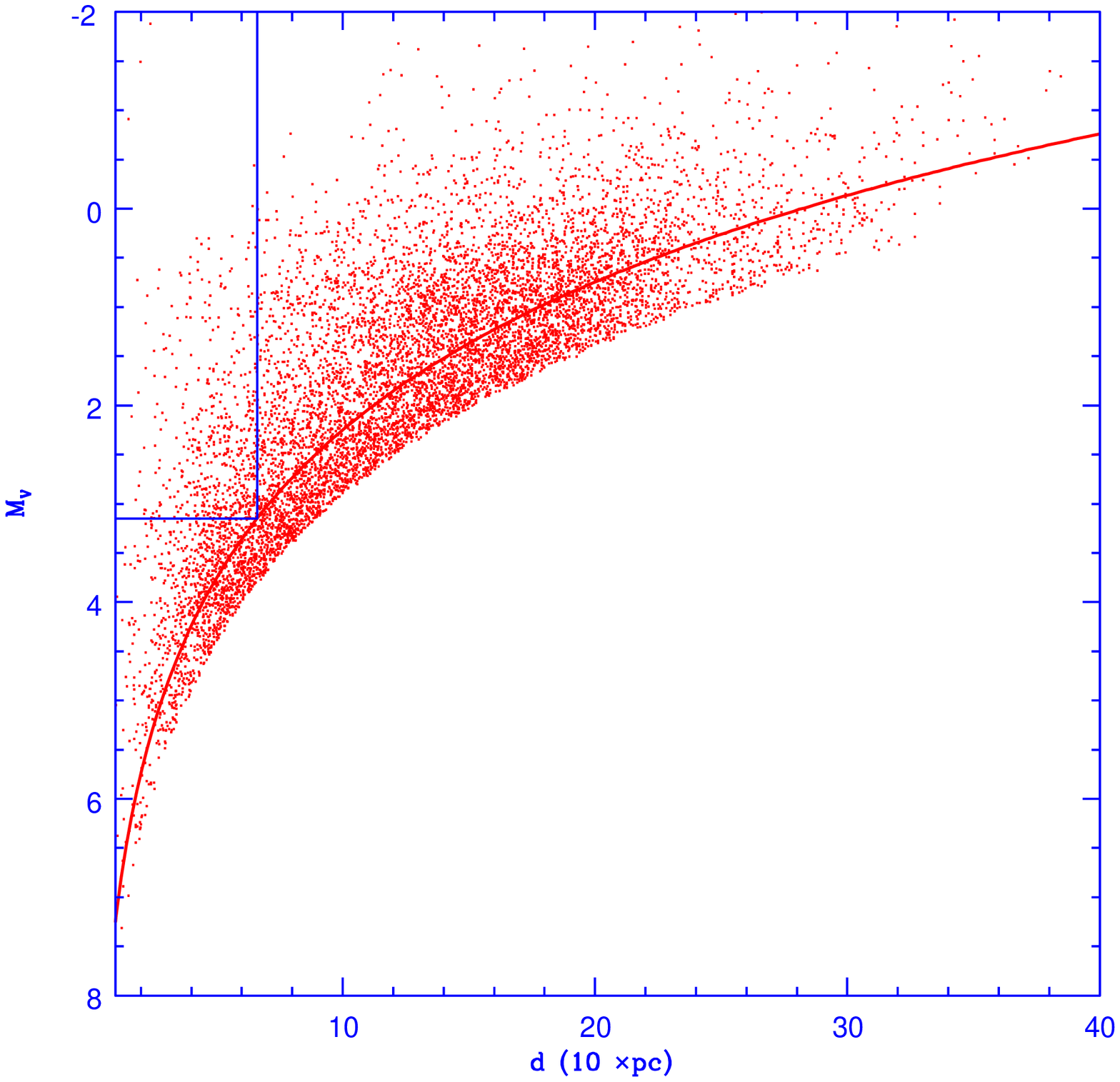,angle=0,width=9.0cm}
\@ \textbf{Figure 3.}\hspace{5pt}
{
Stars in the Hipparcos catalogue having distance errors $< 20$ \%. The sharp
lower envelope shows the completeness limit of m $<7.9$. The straight lines 
show one volume limited sample, at $M_{V} < 3.15$. The curve shows the 
completeness limit of m $<7.25$, which  corresponds to errors similar to 
what was used in the simulations of Figures (1) and (2).
}
\end{figure}

In order to apply the method described in the preceding section to the 
Hipparcos data, we would like to construct a volume-limited sample, where 
no biases appear between stars of  different ages. Further, such a sample 
should contain a sufficient number of stars coming from all age groups 
being considered i.e. it must go down in magnitude to the turnoff point 
corresponding to the oldest age being considered. Although the Hipparcos 
satellite produced a catalogue having very well understood errors and 
highly accurate magnitude and colour determinations for a large number of
stars, the sample has to be reduced through several cuts 
before it complies with the restrictions required by our method.

The Hipparcos catalogue provides an almost complete sample of
stars in the solar neighbourhood. The limiting magnitude depends
both on spectral type and galactic latitude (ESA SP-1200, Volume 1,
page 131). For the types earlier than G5 which we consider here,
the limiting V magnitude is given by $V_{\rm lim} = 7.9 + 1.1 \sin |b|$.
To avoid unnecessary complications, we consider cuts of the type $V$=const.
at all latitudes with $V<7.9$

Figure (3) shows a graph of $M_V$ vs. distance for all  stars in the 
Hipparcos catalogue having distance errors smaller than $20 \%$, and an 
apparent magnitude $m_{V}<7.9$. The solid lines show  the inclusion criteria
for one possible volume limited sample, complete to $M_V < 3.15, m<7.25$ . 
As it can be seen, the maximum age which  can be considered will not be very
large, as the number of stars in a volume-limited sample complete to absolute
magnitudes greater than 4 rapidly dwindles. After experimenting with 
synthetic CMDs of known $SFR(t)$ produced using our isochrone grid and 
constructed to have the same numbers of stars as a function of lower 
magnitude limit as in Figure (3), and recovering the $SFR(t)$ using our 
method,  we identified 3 Gyr as the maximum age we can accurately treat 
with the data at hand. This fixes $t_{0}=0, t_{1}=3$ Gyr as the temporal
limits in equation (2), were the use of 200 isochrones establishes the 
formal resolution of the method to be $15$ Myr.

Although the absolute magnitude errors correlate tightly with the distance, 
the colour errors correlate more strongly with the apparent magnitude, and 
can actually represent the dominant error in inverting the CMDs. The solid 
curve shows the $m_{V}<7.25$ completion limit, which implies errors similar
to those used in Figures (1) and (2). It will be with complete volume-limited
samples having this apparent magnitude limit that we will be dealing.

\begin{figure*}
\epsfig{file=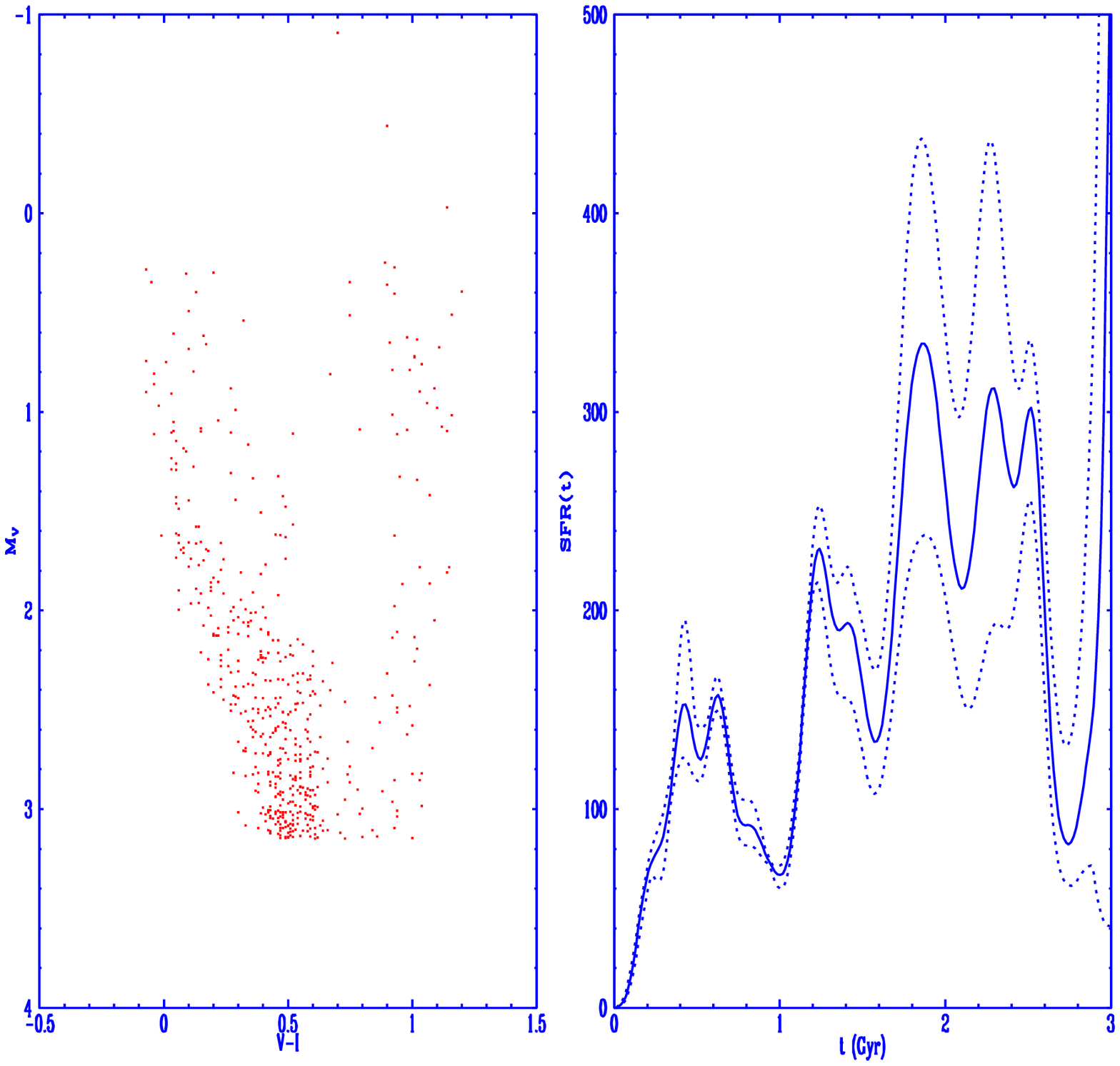,angle=0,width=18.1cm,height=9.0cm}
\@ \textbf{Figure 4.}\hspace{5pt}{
\begin{flushleft}\textbf{Left:} Colour-magnitude diagram of the volume-limited
sample complete to $M_{V} <3.15 $ from the Hipparcos satellite.
\textbf{Right:} Inferred $SFR(t)$ resulting from this dataset. Only stars
bluewards of $V-I=0.7$ where considered in the inversion procedure, 
representative of the last 3 Gyr.
\end{flushleft}}
\end{figure*}

As the limit in $M_{V}$ is moved to dimmer stars, the structure of the $SFR(t)$at older ages is better recovered by the inversion method, but the number of
younger stars diminishes (see Figure 1) and the $SFR(t)$ of the younger 
period is under represented in the recovered  $SFR(t)$. We constructed a 
variety of somewhat independent Hipparcos CMDs for different absolute 
magnitude limits in the range $3.0<M_{V}<3.5$, and obtained highly compatible
answers.

Also, all our samples include a certain number of contaminating stars having
ages greater than the 3 Gyr limit considered by the inversion method, mostly in
the RGB region, as  their turn off points appear at magnitudes dimmer than
the minimum ones considered. To avoid these stars, the inversion method 
considers only stars bluewards of $V-I=0.7$, as was done in the synthetic 
examples of Figures (1) and (2).

As the fraction of stars produced by the $SFR(t)$ which live into the observed
CMD diminishes with age, in inverting a well populated CMD the older regions
of the $SFR(t)$ are under estimated by the recovery method. This is 
compensated by a correction factor given by the assumed IMF, and the mass 
at the tip of the RGB as a function of time, as discussed in paper I.

Once the IMF, metallicity, positions of the observed stars in the CMD and 
observational errors in both coordinates (also supplied  by the Hipparcos 
catalogue) are given, they are used to construct the likelihood matrix 
$G_{i}(t)$, which is the only input given to the inversion method. The small
number of stars present in any volume-limited sample ($\sim 450$) leaves us 
insensitive to the existence of small features in the $SFR(t)$ producing only
a few stars. The limited numbers of stars also reduces the resolution of the 
method, as a smoothing function has to be applied to suppress instabilities 
in solving the integro-differential equation of the problem. This final 
smoothing reduces the effective temporal resolution of the  method to 
50 Myr, still much higher than that of any indirect chemical evolution 
inference of $SFR(t)$.

\subsection{Kinematic and geometric corrections}

Once the apparent and absolute magnitudes of the sample have been chosen, 
the set of stars to be studied is fully specified. The positions of which in
the CMD are compared to the assumed isochrones to construct the likelihood 
matrix, which is the only input required by the numerical implementation of 
the method, as described above. The resulting $SFR(t)$ will be representative
of the stars which where used in the construction of the likelihood matrix.

To normalize the various inferred $SFRs(t)$ from samples having different 
$M_{V}$ limits,  and hence complete out to different distances, we apply the
following kinematic and  geometric corrections.

Let $F(v,h)$ be the fraction of the time a star having vertical velocity at 
the disk plane $v$ spends between heights $-h$ and $+h$. Then, for a 
cylindrical sample complete to height $h$ above and below the disk plane,

\begin{equation}
N(t)={N_{o}(t) \over {\sqrt{2\pi} \, \sigma(t)}} \int_{-\infty} ^{\infty} 
{e^{-v^{2}/{2\sigma(t)^2}} \over F(v,h)} dv
\end{equation}

where $N(t)$ is the number of stars a stellar population of age $t$ contains,
$N_{o}(t)$ the number of stars of age $t$ observed and $\sigma(t)$ the time
dependent velocity dispersion of the several populations. As volume-limited 
samples are generally spherical around the Sun, a further geometric factor 
is required, giving:

\begin{equation}
N(t)={3 N_{o}(t) \over {2\sqrt{2\pi} \, \sigma(t) \,  R^3}} 
\int_{0} ^{R} r \hspace{2pt} h \int_{-\infty} ^{\infty} 
{e^{-v^{2}/{2\sigma(t)^2}}  \over F(v,h)} \, dv \hspace{5pt} dr
\end{equation} 

where $R$ is the radius of the observed spherical volume limited sample, 
$r$ is a radial coordinate and $h^2=R^2-r^2$. To estimate $F(v,h)$ one 
requires the detailed vertical force law of the galactic disk at the solar
neighbourhood. The best direct estimate of this function remains that of 
Kuijken and Gilmore (1989), who show this function to deviate from
that of a harmonic potential to a large degree. This detailed force law 
we integrate numerically to obtain $F(v,h)$. We use 
$\sigma(t) \; = \; 20 \; {\rm km/s} $
which is appropriate for the metallicity and age ranges we are studying 
(Edvardsson et al. 1993, Wyse \& Gilmore 1995). Note also that the
scatter in metallicity within 80 pc is rather small (Garnett and Kobulnicky,
1999) and will not change significantly our results.

In this way, assuming a Gaussian distribution for the vertical
velocities of the stars, and a given $\sigma(t)$, an observed $N_{o}(t)$ 
can be transformed into a total $N(t)$, which is equal to the total 
projected disk quantity. 

In our case the $SFR(t)$ given by the method takes the place of $N_{o}(t)$,
and equation (6) is used to obtain a final star formation history, which 
accounts for the kinematic and geometric factors described. This function 
is then normalized through the total number of stars in the relevant sample,
to give the deduced $SFR(t)$ in units of $M_{\odot}$Myr$^{-1}$kpc$^{-2}$.

\subsection{Results}

Figure (4) shows the CMD corresponding to a volume-limited sample complete
to $M_{V}<3.15$ for stars in the Hipparcos catalogue having errors in 
parallax of less than $20 \%$ and  $m_{V}<7.25$ (left panel). The right panel
of this figure shows the result of applying our inversion method to this CMD,
solid curve. The dotted envelope encloses several alternative reconstructions
arising from different $M_{V}$ cuts, and gives an estimate of the errors 
likely to be present in our result, which can be seen to increase with time. 
The reconstruction based on the $(M_V, B-V)$ diagram gives essentially the
same results.

\begin{figure*}
\epsfig{file=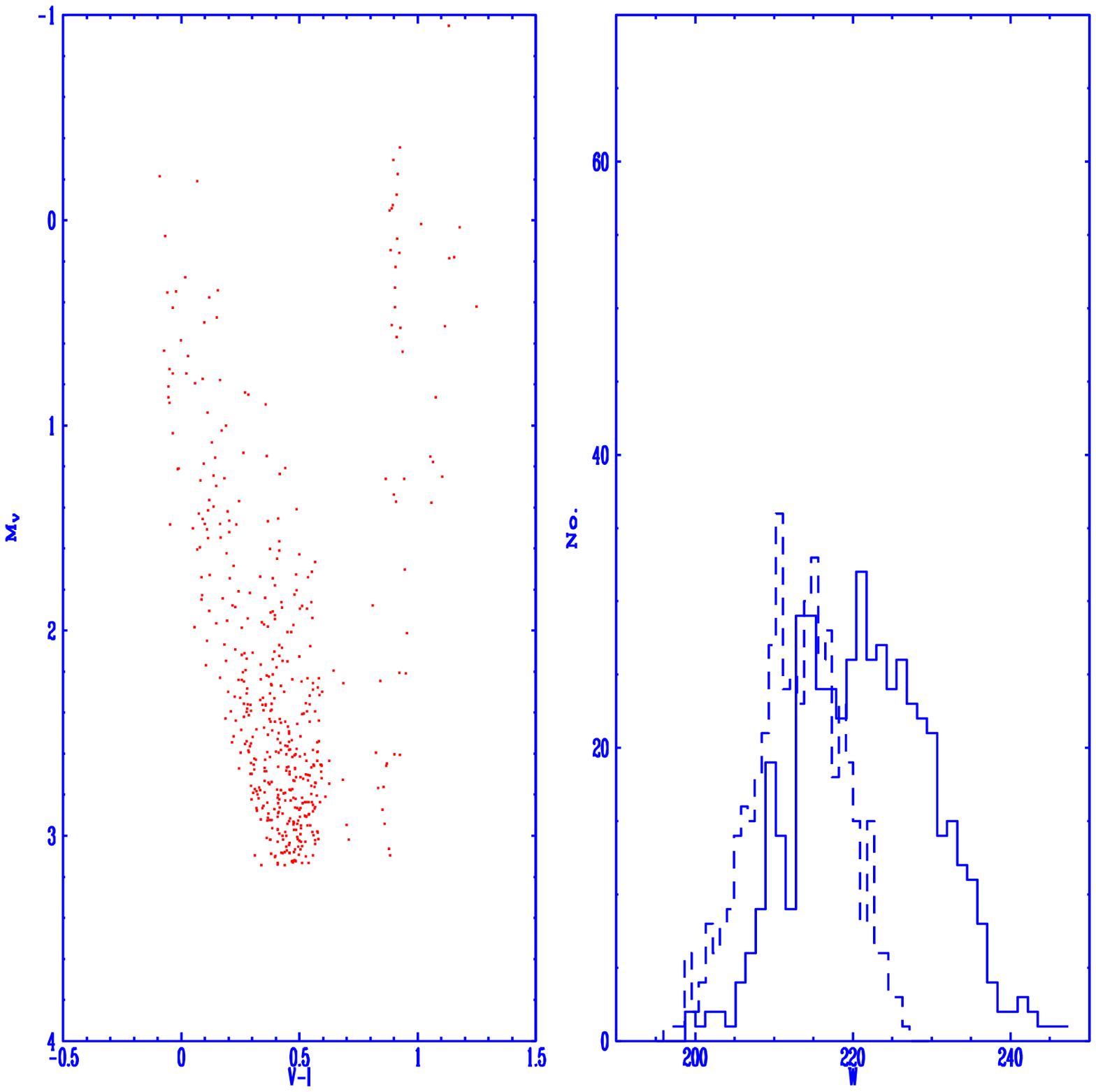,angle=0,width=18.1cm,height=9.0cm}
\@ \textbf{Figure 5.}\hspace{5pt}{
\begin{flushleft}
\textbf{Left:} Simulated CMD from the inferred $SFR(t)$, down to $M_{V} <3.15$
containing a similar number of stars to the volume limited sample complete to
the same limit.
\textbf{Right:} Histogram of W statistics for 500 model-model comparisons, 
solid curve. The dashed histogram gives the W statistics of 500 data-model
comparisons, showing the inferred $SFR(t)$ to be compatible with this 
volume-limited sample.
\end{flushleft}}
\end{figure*} 

A certain level of constant star formation activity can be seen, superimposed
onto a strong, quasi-periodic component having a period close to $0.5$ Gyr, as 
encoded in the positions of the observed stars in the CMD. The sharp feature 
seen towards $t=3$ Gyr could be the beginning of a fifth cycle, truncated by 
the boundary condition  $SFR(3)=0$. We have performed tests with synthetic 
CMDs having the same numbers of stars and magnitude limits as in Figure (4), 
and having a variety of $SFR(t)$. The method efficiently discriminates 
between constant and periodic input $SFR(t)$, and correctly recovers features
such as those found in the inferred $SFR(t)$ of Figure (4). We conclude that
as far as the Padova isochrones at solar metallicity are representative of the
observational properties of the stars in the CMDs, the $SFR(t)$ of the
solar neighbourhood over the last 3 Gyr has been that shown in Figure (4). 
The unprecedented time resolution of our $SFR$ reconstruction makes it
difficult to compare with the results derived from chromospheric
activity studies (Rocha-Pinto et al. 1999), although qualitatively we
do find the same activity at both 0.5 and above 2 Gyr, but not the
decrease between 1 and 2 Gyr.

One possible interpretation of a cyclic component in the $SFR(t)$ of the 
solar neighbourhood can be found in the density wave hypothesis (Lin and Shu,
 1964) for the presence of spiral arms in late type galaxies. As the pattern
speed and the circular velocity are in general different, a given region of 
the disk (e.g. the solar neighbourhood), periodically crosses an 
arm region where the increased local gravitational potential might possibly
trigger an episode  of star formation. In the simplest version of this 
scenario, we can take  the pattern angular frequency $\Omega_p$ 
equal to twice the circular frequency $\Omega$ at the Sun's 
position (Binney \& Tremaine 1987),
valid  within a flat rotation curve region. The time interval $\Delta t$ 
between 
encounters with an arm at the solar neighbourhood will be given in general by 
$$
\Delta t \; = \; \frac{2 \, \pi}{m \, | \Omega - \Omega_p |}
$$
that is, 
$$
\Delta t \; = \; \frac{0.22 \, {\rm Gyr}}{m} \; \left(\frac{\Omega}{29 \,
 {\rm km \, s}^{-1} {\rm kpc}^{-1}}\right)^{-1} \; 
 \left| \frac{\Omega_p}{\Omega} - 1  \right|^{-1}
$$
 
where $m$ is the number of arms in the spiral pattern. The classical
value of the pattern speed, $\Omega_p = 0.5 \, \Omega \approx 14.5 $ 
 km s$^{-1}$ kpc$^{-1}$ would imply that
the interaction with a single arm ($m=1$) would be enough to account
for the observed regularity in the recent SFR history. However, more recent
determinations tend to point to much larger values (e.g.  
Mishkurov et al. 1979, Avedisova 1989, Amaral and L\'epine 1997) close to
$\Omega_p \sim 23 - 24$ km s$^{-1}$ kpc$^{-1}$, which would then imply that 
the regularity
present in the reconstructed $SFR(t)$ would be consistent with a scenario
where the interaction of the solar neighbourhood with a two-armed spiral
pattern would have induced the star formation episodes we detect. These
arms are clearly detected in, for instance, the distribution of free
electrons in the galactic plane (Taylor and Cordes 1993). This
is reminiscent of the explanations put forward to account for the
inhomogeneities observed in the velocity distribution function, where
well-defined branches associated with moving groups of different
ages (Chereul et al. 1999, Skuljan et al. 1999, Asiain et al. 1999) 
could perhaps be also associated with an interaction with spiral arm(s),
although in this case the time scales are much smaller.

Alternatively, if the solar neighbourhood is closer to the corotation radius,
the galactic bar could have triggered star formation in the solar
neighbourhood with episodes separated by about 0.5 Gyr if the pattern
speed of the bar is larger than about 40 km s$^{-1}$ kpc$^{-1}$ (Dehnen 1999).

Of course, other explanations are possible, for example
the cloud formation, collision and stellar feedback models of Vazquez \& 
Scalo (1989) predict a phase of oscillatory $SFR(t)$ behaviour as a result 
of a self-regulated star formation r\'egime. Close encounters with the
Magellanic Clouds have also been suggested to explain the intermittent  
nature of the SFR on longer time scales (Rocha-Pinto et al. 2000).

We have tested the ability of our method to accurately distinguish oscillatory components in the
$SFR(t)$ with tests such as those shown in Figures (1) and (2). The oscillatory component
in the case shown in Figure (1) is successfully recovered by the method. In the case
shown in Figure(2) however, although the main features are also accurately recovered, a level 
of small amplitude fluctuations spuriously appears. This last feature is of such a small
level, that if a CMD where produced from the method answer of Figure (1) having the same total
number of stars, each small fluctuation would produce a number of stars of order 1.

Our answer shown in Figure (4) shows not only a large scale oscillatory component, but
superimposed onto this, a certain level of small amplitude fluctuations.
Given the total number of stars present in our sample, we can not rule out the 
possibility (quit possibly  in fact the case) that these small fluctuations are 
numerical, as they are of amplitude similar to the ones discussed appearing in Figure(2), 
and are actually buried within the error envelope. The main oscillatory component having
a period of 0.5 Gyr however, involves a sufficiently large number of stars to be objectively 
identified.

A larger and independent
data set from which to derive $SFR(t)$ would be necessary in order to extend
our results to a broader age range, and a more extensive region of the Galactic disk.
Increasing the number of stars available for the HR inversion procedure would also allow
to recover finer features, and reduce the small numerical fluctuations discussed above.

\section{Testing the results}

In our Paper I we tested this method using  synthetic CMDs produced from 
known star formation histories, with which we could assess the accuracy of the
result
of the inversion procedure, as shown in Figures (1) and (2). In working with
real data, we require the introduction of an independent method of comparing
our final result to the starting CMD, in order to check that the answer our
inversion procedure gives is a good answer. From our paper I we know that 
when the stars being used in the inversion procedure were indeed produced 
from the isochrones and metallicity used to construct the likelihood matrix,
the inversion method gives accurate results. The introduction of  an 
independent comparison between our answer and the data is hence a way of 
checking the accuracy  of the input physics used in the inversion procedure,
 i.e. the IMF, metallicity and observational parameters.

The most common procedure of comparing a certain $SFR(t)$ with an observed 
CMD is to use the $SFR(t)$ to generate a synthetic CMD, and compare this 
to the observations using a statistical test to determine the degree of 
similarity between the two.

The disadvantage however is that one is not comparing the $SFR(t)$ with 
the data, but rather a particular realisation of the $SFR(t)$ with the data.
The distinction becomes arbitrary when large numbers of stars are found 
in all regions of the CMD, which is generally not the case. Following a 
Bayesian approach, we prefer to adopt the $W$ statistic presented by 
Saha (1998), essentially

$$
W=\prod_{i=1}^{B} {{(m_{i}+s_{i})!} \over {m_{i}!s_{i}!}}
$$

where B is the number of cells into which the CMD is split, and $m_{i}$ 
and $s_{i}$ are the numbers of points two distributions being compared have
in each cell. This asks for the probability that two distinct data sets 
are random realisations of the same underling distribution. In  implementing 
this test we first produce a large number (500) of random realisations of our 
inferred $SFR(t)$, and compute the $W$ statistic between pairs in this 
sample of CMD's. This gives a distribution which is used to determine a 
range of values of $W$ which are expected to arise in random realisations 
of the $SFR(t)$ being tested. Next the $W$ statistic is computed between 
the observed data set, and a new large number of random realisations of 
$SFR(t)$ (also 500), this gives a new distribution of $W$ which can be 
objectively compared to the one arising from the model-model comparison 
to assess whether both data and modeled CMD's are compatible with a unique 
underling distribution.

Figure (5) shows a synthetic CMD produced from our inferred $SFR(t)$ for 
the solar neighbourhood, down to $M_{V}=3.15$. This can be compared to 
the Hipparcos CMD complete to the same $M_{V}$ limit of Figure (4). A 
visual inspection reveals approximately equal numbers of stars in each of 
the distinct regions of the diagram, a more rigorous statistical comparison
is also included. The right panel of Figure (5) shows a histogram of the 
values of the $W$ statistic for 500 random realisations of our inferred 
$SFR(t)$ in a model-model comparison. This gives the range of values of 
the $W$ statistic likely to appear in comparisons of two CMD diagrams 
arising from the same underlying $SFR(t)$, our inferred $SFR(t)$. The dashed
histogram shows the results of 500 synthetic CMDs vs. the observed Hipparcos
data set. The compatibility of both sets of $W$ values is clear. In this 
way the observed stars in the volume limited sample complete to $M_{V}=3.15$
and the isochrones, IMF and metallicity used in estimating the inferred 
$SFR(t)$ shown in Figure (4) are shown to be compatible with each other. 
Similar tests were also performed changing the limiting $M_{V}$ in the
range $3.0-3.5$, and comparing against the corresponding Hipparcos sample. 
This produces alternative CMDs which contain different stars (see Figure 3), 
which were compared to synthetic CMDs coming always from the same central 
inferred $SFR(t)$. The results were always equal or better than what is
shown in Figure 5, model-model and data-model distributions of $W$ having 
mean values well within $1 \sigma$ of each other, where $\sigma$ refers 
to either the model-model or the data-model $W$ distributions.

\section{Conclusion}

We have applied the method developed in our paper I to the data of the 
Hipparcos catalogue. An objective answer for the $SFR(t)$ of the solar 
neighbourhood over the last 3 Gyr was found, which can be shown to be 
consistent with the complete volume-limited Hipparcos samples relevant 
to this age range. A structured $SFR(t)$ is obtained showing a cyclic 
pattern having a period of about 0.5 Gyr, superimposed on some degree of 
underlying star formation activity which increases slightly with age. 
No random bursting behaviour was found at the time resolution of 0.05 
Gyr of our method. A first order density wave  model for the repeated 
encounter of galactic arms could explain the observed regularity.

\end{document}